\documentclass[twoside,reqno]{bjp}
\usepackage{epsfig,cite}
\usepackage{graphics}
\usepackage{amssymb,amsmath}
\usepackage{times}
\begin{document}

\sloppy \raggedbottom

 \setcounter{page}{1}


% use the thanks command within \title, \author or \address for footnotes;
% \title{Title} or  \title{Title\thanks{...}}
% \author{Name1}{aff.label1}, \coauthor{Name2}{aff.label2},  \coauthor{Name3}{aff.label3}
% \address{Address1}{aff.label1}
% \address{Address2}{aff.label2}
% \address{Address3}{aff.label3}
%\runningheads{NAMES OF AUTHORS IN CAPITALS}{SHORT TITLE IN CAPITALS}

\title{Advance of Planetary Perihelion in Post-Newtonian Gravity}

\runningheads{Advance of Planetary Perihelion in Post-Newtonian Gravity}{A. Larra\~{n}aga,
L. Cabarique}

\begin{start}
\author{Alexis Larra\~{n}aga}{1}, \coauthor{Luis Cabarique}{2}

\address{National Astronomical Observatory. National University of Colombia.}{1}

\address{Department of Physics. National University of Colombia.}{2}

\received{23 August 2002}

\begin{Abstract}
We present an elementary derivation of the planetary advance of the
perihelion for a general spherically symmetric line element in the
post-newtonian approximation.
\end{Abstract}

\end{start}

\section[]{Introduction}

Einstein field equations are nonlinear, and therefore cannot in general
be solved exactly. By imposing the symmetry requirements of time independece
and spatial isotropy, the Schwarzschild solution can be obtained,
but this solution can not actually describe in the solar system because
it is not static and isotropic. The Post-Newtonian approximation is
a systematic method in which we can describe a system of slowly moving
particles bound together by gravitational forces. In this method we
will use an expansion of geometric objects as the metric tensor, in
inverse powers of the speed of light using the parameter $\beta=\frac{v}{c}$.

Several works had calculated the perihelion precession of planetary
orbits based on Einstein's equations with the line element of Schwarzschild
and the use of Euler-Lagrange equations. In this paper we present
a simplified derivation using a post-newtonian approximation for a
general spherically symmetric line element. In order to obtain the
precession \cite{cornbleet93}, we compare the Keplerian orbit with
the curved space trajectory and making use of the invariance of Kepler's
second law.

\section{Post-Newtonian Metric}

The Minkowski metric in polar coordinates is given by

\begin{equation}
ds^{2}=dt^{2}-dr^{2}-r^{2}d\vartheta^{2}-r^{2}\sin^{2}\vartheta\, d\varphi^{2}\label{ds-plano}
\end{equation}

whereas a general curved metric with spherical symmetry has a line
element that can be written as

\begin{equation}
ds^{2}=g_{tt}(r)dt^{2}+g_{rr}(r)dr^{2}-r^{2}d\vartheta^{2}-r^{2}\sin^{2}\vartheta d\varphi^{2}.\label{ds-gral}
\end{equation}

In the post-newtonian approximation the metric components $g_{tt}$
and $g_{rr}$ are expanded up to the fourth order in $\beta=\frac{v}{c}$
in the form

\begin{eqnarray}
g_{tt} & = & 1+\overset{\left(2\right)}{g}_{tt}+\overset{\left(4\right)}{g}_{tt}+O\left(\beta^{6}\right)\\
g_{rr} & = & -1+\overset{\left(2\right)}{g}_{rr}+\overset{\left(4\right)}{g}_{rr}+O\left(\beta^{6}\right).
\end{eqnarray}

Thus, the relation between flat and curved spaces for the radial and
time coordinates can be considered as the transformation

\begin{equation}
dt'=\left[1+\frac{1}{2}\overset{\left(2\right)}{g}_{tt}+\frac{1}{2}\overset{\left(4\right)}{g}_{tt}\right]dt\label{trans-t}
\end{equation}
 and 
\begin{equation}
dr'=\left[1-\frac{1}{2}\overset{\left(2\right)}{g}_{rr}-\frac{1}{2}\overset{\left(4\right)}{g}_{rr}\right]dr,\label{trans-r}
\end{equation}
where the binomial approximation has been applied.

\section{The Advance of the Perihelion }

In order to obtain the advance of the perihelion we will consider
two elliptical orbits. First, the Keplerian orbit in flat space with
coordinates $\left(t,r\right)$ for which the element of area is given
by

\begin{equation}
dA=\int_{0}^{R}rdrd\vartheta=\frac{R^{2}}{2}d\vartheta.
\end{equation}
This equation gives the second law of Kepler,

\begin{equation}
\frac{dA}{dt}=\frac{R^{2}}{2}\frac{d\vartheta}{dt}.
\end{equation}

The second elliptical orbit that we will consider is obtained in the
coordinates $\left(t',r'\right)$ of the curved space. Here, the element
of area is

\begin{eqnarray}
dA' & = & \int_{0}^{R}rdr'd\vartheta\\
dA' & = & \int_{0}^{R}r\left[1-\frac{1}{2}\overset{\left(2\right)}{g}_{rr}-\frac{1}{2}\overset{\left(4\right)}{g}_{rr}\right]drd\vartheta\\
dA' & = & \dfrac{R^{2}}{2}\left(1-\dfrac{1}{R^{2}}\int_{0}^{R}r\left[\overset{\left(2\right)}{g}_{rr}+\overset{\left(4\right)}{g}_{rr}\right]dr\right)d\vartheta
\end{eqnarray}

where we have used (\ref{trans-r}). Kepler's second law is now

\begin{equation}
\frac{dA'}{dt'}=\dfrac{R^{2}}{2}\left(1-\dfrac{1}{R^{2}}\int_{0}^{R}r\left[\overset{\left(2\right)}{g}_{rr}+\overset{\left(4\right)}{g}_{rr}\right]dr\right)\frac{d\vartheta}{dt'},
\end{equation}

which, using equation (\ref{trans-t}), gives

\begin{equation}
\frac{dA'}{dt'}=\dfrac{R^{2}}{2}\left(1-\dfrac{1}{R^{2}}\int_{0}^{R}r\left[\overset{\left(2\right)}{g}_{rr}\left(r\right)+\overset{\left(4\right)}{g}_{rr}\left(r\right)\right]dr\right)\left[1+\frac{1}{2}\overset{\left(2\right)}{g}_{tt}\left(R\right)+\frac{1}{2}\overset{\left(4\right)}{g}_{tt}\left(R\right)\right]^{-1}\frac{d\vartheta}{dt}.
\end{equation}
Using again the binomial approximation it gives

\begin{equation}
\frac{dA'}{dt'}=\dfrac{R^{2}}{2}\Phi\left(R\right)\frac{d\vartheta}{dt}
\end{equation}

where we have defined

\begin{equation}
\Phi\left(R\right)=\left(1-\dfrac{1}{R^{2}}\int_{0}^{R}r\left[\overset{\left(2\right)}{g}_{rr}\left(r\right)+\overset{\left(4\right)}{g}_{rr}\left(r\right)\right]dr\right)\left[1-\frac{1}{2}\overset{\left(2\right)}{g}_{tt}\left(R\right)-\frac{1}{2}\overset{\left(4\right)}{g}_{tt}\left(R\right)\right].
\end{equation}

Applying this increase to change $d\vartheta$ to $d\vartheta'$ in
a single orbit gives the expression

\begin{equation}
\int_{0}^{\Delta\vartheta'}d\vartheta'=\Delta\vartheta'=\int_{0}^{\Delta\vartheta=2\pi}\Phi(R)d\vartheta.
\end{equation}

In order to perform the integration in the right hand side we will
consider the Keplerian ellipse using $R\left(\vartheta\right)=\frac{l}{1+\epsilon\cos\vartheta}$
where $l$ is the \emph{latus rectum }and $\epsilon$ is the eccentricity.

\section{Particular Cases}

\subsection{Schwarzschild Metric}

Up to second order in $\beta^{2}$, Schwarzschild's metric is

\begin{equation}
ds^{2}=\left(1-\frac{2M}{r}\right)dt^{2}-\left(1+\frac{2M}{r}\right)dr^{2}-r^{2}d\vartheta^{2}-r^{2}\sin^{2}\vartheta d\varphi^{2},
\end{equation}
and therefore

\begin{equation}
\overset{\left(2\right)}{g}_{rr}\left(r\right)=\overset{\left(2\right)}{g}_{tt}\left(r\right)=-\frac{2M}{r}
\end{equation}

\begin{equation}
\overset{\left(4\right)}{g}_{rr}\left(r\right)=\overset{\left(4\right)}{g}_{tt}\left(r\right)=0.
\end{equation}

Function $\Phi\left(R\right)$ is given by

\begin{eqnarray}
\Phi\left(R\right) & = & \left(1-\dfrac{1}{R^{2}}\int_{0}^{R}r\overset{\left(2\right)}{g}_{rr}\left(r\right)dr\right)\left[1-\frac{1}{2}\overset{\left(2\right)}{g}_{tt}\left(R\right)\right]\\
\Phi\left(R\right) & = & \left(1+\dfrac{1}{R^{2}}\int_{0}^{R}2Mdr\right)\left(1+\frac{M}{R}\right)\\
\Phi\left(R\right) & = & \left(1+\dfrac{2M}{R}\right)\left(1+\frac{M}{R}\right)\\
\Phi\left(R\right) & = & 1+\frac{3M}{R}+\frac{2M^{2}}{R^{2}}
\end{eqnarray}

and therefore the increasing $\Delta\vartheta'$ is 

\begin{equation}
\Delta\vartheta'=\int_{0}^{\Delta\vartheta=2\pi}\left[1+\frac{3M}{R}+\frac{2M^{2}}{R^{2}}\right]d\vartheta.
\end{equation}
In order to perform the integration, we will the Keplerian ellipse
using $R\left(\vartheta\right)=\frac{l}{1+\epsilon\cos\vartheta}$,
so

\begin{equation}
\Delta\vartheta'=\int_{0}^{\Delta\vartheta=2\pi}\left[1+\frac{3M}{l}\left(1+\epsilon\cos\vartheta\right)+\frac{2M^{2}}{l^{2}}\left(1+\epsilon\cos\vartheta\right)^{2}\right]d\vartheta
\end{equation}

\begin{equation}
\Delta\vartheta'=2\pi+\frac{6\pi M}{l}+\frac{2\pi M^{2}}{l^{2}}\left(2+\epsilon^{2}\right).
\end{equation}

Hence, the obtained perihelion advance for Schwarzschild's metric
has the standard value $\frac{6\pi M}{l}$ plus an additional term
of order $\beta^{4}$.

\subsection{Reissner-Nordström Metric}

Up to fourth order in $\beta$, the metric for an electrically charged
and spherically symmetric object is

\begin{equation}
ds^{2}=\left(1-\frac{2M}{r}+\frac{Q^{2}}{r^{2}}\right)dt^{2}-\left(1+\frac{2M}{r}-\frac{Q^{2}}{r^{2}}\right)dr^{2}-r^{2}d\vartheta^{2}-r^{2}\sin^{2}\vartheta d\varphi^{2},
\end{equation}
and therefore we will take the approximation terms

\begin{equation}
\overset{\left(2\right)}{g}_{rr}\left(r\right)=\overset{\left(2\right)}{g}_{tt}\left(r\right)=-\frac{2M}{r}
\end{equation}
\\
\begin{equation}
\overset{\left(4\right)}{g}_{tt}\left(r\right)=\frac{Q^{2}}{r^{2}}.
\end{equation}

This time function $\Phi\left(R\right)$ is 

\begin{eqnarray}
\Phi\left(R\right) & = & \left(1-\dfrac{1}{R^{2}}\int_{0}^{R}r\overset{\left(2\right)}{g}_{rr}\left(r\right)dr\right)\left[1-\frac{1}{2}\overset{\left(2\right)}{g}_{tt}\left(R\right)-\frac{1}{2}\overset{\left(4\right)}{g}_{tt}\left(R\right)\right]\\
\Phi\left(R\right) & = & \left(1+\dfrac{1}{R^{2}}\int_{0}^{R}2Mdr\right)\left(1+\frac{M}{R}-\frac{Q^{2}}{2R^{2}}\right)\\
\Phi\left(R\right) & = & \left(1+\dfrac{2M}{R}\right)\left(1+\frac{M}{R}-\frac{Q^{2}}{2R^{2}}\right)\\
\Phi\left(R\right) & = & 1+\frac{3M}{R}+\frac{2M^{2}}{R^{2}}-\frac{Q^{2}}{2R^{2}}-\frac{MQ^{2}}{R^{3}}
\end{eqnarray}

and the increasing $\Delta\vartheta'$ is now

\begin{equation}
\Delta\vartheta'=\int_{0}^{\Delta\vartheta=2\pi}\left[1+\frac{3M}{R}+\frac{2M^{2}}{R^{2}}-\frac{Q^{2}}{2R^{2}}-\frac{MQ^{2}}{R^{3}}\right]d\vartheta.
\end{equation}
In order to perform the integration we consider again $R\left(\vartheta\right)=\frac{l}{1+\epsilon\cos\vartheta}$,
and therefore

\begin{equation}
\Delta\vartheta'=\int_{0}^{\Delta\vartheta=2\pi}\left[1+\frac{3M}{l}\left(1+\epsilon\cos\vartheta\right)+\left(\frac{2M^{2}}{l^{2}}-\frac{Q^{2}}{2l^{2}}\right)\left(1+\epsilon\cos\vartheta\right)^{2}-\frac{MQ^{2}}{l^{3}}\left(1+\epsilon\cos\vartheta\right)^{3}\right]d\vartheta
\end{equation}

\begin{equation}
\Delta\vartheta'=2\pi+\frac{6\pi M}{l}+\frac{2\pi M^{2}}{l^{2}}\left(2+\epsilon^{2}\right)-\frac{\pi Q^{2}}{2l^{2}}\left(2+\epsilon^{2}\right)-\frac{\pi MQ^{2}}{l^{3}}\left(2+3\epsilon^{2}\right)
\end{equation}

We obtained again the standard value $\frac{6\pi M}{l}$ and the $\beta^{2}$
order term for Schwarzschild's solution given above, but now we also
obtain an additional term of order $\beta^{4}$ solely due to the
electric charge of the central body which agree with the perihelion
precession reported in \cite{chaliasos01} and \cite{gong09}. Finally
we obtain a term that is proportional to the product $MQ^{2}$ that
is not reported in those papers.

Finally, it is also important to note that the terms that contain
the electric charge have a negative contribution to the perihelion
precession.

\subsection{Schwarzschild-de Sitter and Schwarzschild-Anti de Sitter Metric}

The Schwarzschild-de Sitter metric is

\begin{equation}
ds^{2}=\left(1-\frac{2M}{r}+\Lambda r^{2}\right)dt^{2}-\left(1+\frac{2M}{r}+\Lambda r^{2}\right)dr^{2}-r^{2}d\vartheta^{2}-r^{2}\sin^{2}\vartheta d\varphi^{2}
\end{equation}

where $\Lambda$ is the cosmological constant and therefore we will
take

\begin{equation}
\overset{\left(2\right)}{g}_{rr}\left(r\right)=\overset{\left(2\right)}{g}_{tt}\left(r\right)=-\frac{2M}{r}
\end{equation}
\\
\begin{equation}
\overset{\left(4\right)}{g}_{rr}\left(r\right)=\overset{\left(4\right)}{g}_{tt}\left(r\right)=\Lambda r^{2}.
\end{equation}

This time function $\Phi\left(R\right)$ is 

\begin{eqnarray}
\Phi\left(R\right) & = & \left(1-\dfrac{1}{R^{2}}\int_{0}^{R}r\left[\overset{\left(2\right)}{g}_{rr}\left(r\right)+\overset{\left(4\right)}{g}_{rr}\left(r\right)\right]dr\right)\left[1-\frac{1}{2}\overset{\left(2\right)}{g}_{tt}\left(R\right)-\frac{1}{2}\overset{\left(4\right)}{g}_{tt}\left(R\right)\right]\\
\Phi\left(R\right) & = & \left(1+\dfrac{1}{R^{2}}\int_{0}^{R}\left[2M-\Lambda r^{3}\right]dr\right)\left(1+\frac{M}{R}-\frac{\Lambda R^{2}}{2}\right)\\
\Phi\left(R\right) & = & \left(1+\dfrac{2M}{R}-\frac{1}{4}\Lambda R^{2}\right)\left(1+\frac{M}{R}-\frac{\Lambda R^{2}}{2}\right)\\
\Phi\left(R\right) & = & 1+\frac{3M}{R}+\frac{2M^{2}}{R^{2}}-\frac{5\Lambda MR}{4}-\frac{3\Lambda R^{2}}{4}+\frac{\Lambda^{2}R^{4}}{8}
\end{eqnarray}

and the increasing $\Delta\vartheta'$ is now

\begin{equation}
\Delta\vartheta'=\int_{0}^{\Delta\vartheta=2\pi}\left[1+\frac{3M}{R}+\frac{2M^{2}}{R^{2}}-\frac{5\Lambda MR}{4}-\frac{3\Lambda R^{2}}{4}+\frac{\Lambda^{2}R^{4}}{8}\right]d\vartheta.
\end{equation}
In order to perform the integration we consider again $R\left(\vartheta\right)=\frac{l}{1+\epsilon\cos\vartheta}$,
and therefore

\begin{eqnarray}
\Delta\vartheta' & = & \int_{0}^{\Delta\vartheta=2\pi}\left[1+\frac{3M}{l}\left(1+\epsilon\cos\vartheta\right)+\frac{2M^{2}}{l^{2}}\left(1+\epsilon\cos\vartheta\right)^{2}\right.\nonumber \\
 &  & \left.-\frac{5\Lambda Ml}{4\left(1+\epsilon\cos\vartheta\right)}-\frac{3\Lambda l^{2}}{4\left(1+\epsilon\cos\vartheta\right)^{2}}+\frac{\Lambda^{2}l^{4}}{8\left(1+\epsilon\cos\vartheta\right)^{4}}\right]d\vartheta
\end{eqnarray}

which for small eccentricities can be integrated as

\begin{equation}
\Delta\vartheta'=2\pi+\frac{6\pi M}{l}+\frac{2\pi M^{2}}{l^{2}}\left(2+\epsilon^{2}\right)-\frac{5\pi\Lambda Ml}{2}-\frac{3\pi\Lambda l^{2}}{2}+\frac{\pi\Lambda^{2}l^{4}}{4}
\end{equation}

where the three last terms are the corrections due to the cosmological
constant. The classical advance in the perihelion is recuperated for
zero cosmological constant $\left(\Lambda\rightarrow0\right)$ and
the complete expression agrees with the one reported in \cite{cruz05}.
This derivation is much simpler than the obtained in the work of G.
Kraniotis and S. Whitehouse \cite{kraniotis03}, where the evaluation
was done by means of the genus 2 siegelsche modular forms and including
the Mercury\textquoteright{}s data. \\

\section*{Acknowledgments}

This work was supported by the Universidad Nacional de Colombia. Hermes
Project Code 13038.

\end{document}